\documentclass[11pt,fleqn]{article}
\usepackage{amssymb}

\usepackage{amsmath}

\usepackage{amsmath,amssymb,amsthm}

\begin{document}

\begin{center}

{\Large \textbf{Differential Invariants of Infinite-Dimensional
Algebras That Are Equivalence Algebras of Classes of PDE}}

\vskip 10pt {\large \textbf{Irina YEHORCHENKO}}

\vskip 10pt {Institute of Mathematics of NAS Ukraine, 3
Tereshchenkivs'ka Str., 01601 Kyiv-4, Ukraine} E-mail:
iyegorch@imath.kiev.ua

\vskip 10pt Bulg. J. Phys. vol.35 no.1 (2008), pp. 481-488

\end{center}

\begin{abstract}
We describe differential invariants of infinite-dimensional
algebras being equivalence algebras of some classes of PDE and
study structure of these algebras.
\end{abstract}

\section{Introduction}

We consider infinite-dimensional algebras in the framework of the
applications of the Lie theory and symmetry analysis of
differential equations. We will discuss finding and full
description of differential invariants for such algebras (for
basic concepts on differential invariants of Lie algebras see
\cite{Yehorchenko:LieDI}). The methods and concepts presented here
for infinite-dimensional algebras are based on similar concepts
for finite-dimensional Lie algebras (\cite{Yehorchenko:Ovs-eng,
Yehorchenko:book1,Yehorchenko:book2}). Some aspects of
differential invariants for pseudogroups (another term for
infinite-dimensional counterparts of local Lie transformation
groups) were considered in \cite{Yehorchenko:OlverPohjanpelto}.

Here we are studying a special class of infinite-dimensional -
equivalence algebras of classes of differential equations
(algebras of equivalence transformations, see e.g.
\cite{Yehorchenko:book2}).

As a simple but illustrative example we will be considering a
class of equations
\begin{equation} \label{wave class}
u_{tt} - u_{xx}=f(u, u_t^2-u_x^2),
\end{equation}

\noindent where $u=u(t,x)$,

$$ u_t = \frac{\partial u}{\partial t},
 u_x = \frac{\partial u}{\partial x}, u_{tt}=\frac{\partial^2 u}{\partial
 t^2},u_{xx}=\frac{\partial^2 u}{\partial
 x^2},
$$

\noindent $f$ is an arbitrary function of its arguments (parameter
function of the class of equations).

{\bf Definition 1.} {\it An equivalence transformation for a class
of equations \eqref{wave class} is a change of variables of the
form
\begin{equation} \label{eq-transf}
t'=\tau(t,x,u), x'=\kappa(t,x,u), u'=\omega(t,x,u),
f'=\varphi(t,x,u,f)
\end{equation}
that transforms each equation of the form \eqref{wave class} into
an equation of the same form
\begin{equation} \label{wave class prime}
u'_{t't'} - u'_{x'x'}=f'(u', u^{'2}_{t'}, u^{'2}_{x'}).
\end{equation}
The equations \eqref{wave class} and \eqref{wave class prime} are
said to be equivalent. }

Operators of the equivalence algebra for the class of equation
\eqref{wave class} where found in \cite{Yehorchenko:TTV2} (the
notation $\sigma=u_t^2-u_x^2$ is used):

$$
Y_0=x\partial_t+t\partial_x- u_x\partial_{u_t}-u_t\partial_{u_x},
Y_1=\partial_t, Y_2=\partial_x,
$$
\begin{equation}
Y_3=x\partial_x+t\partial_t- 2f\partial_f- \sigma\partial_\sigma
-2f_u\partial_{f_u},\label{eq algebra}
\end{equation}
$$
 Y_\phi=\phi\partial_u+2\phi'\sigma\partial_\sigma+2(\phi'f+\phi''\sigma)\partial_f
$$
$$
+ (\phi''f+\phi'''\sigma-2\sigma f_\sigma\phi'')\partial_{f_u}+
(\phi''-f_\sigma\phi')\partial_{f_\sigma},
$$

\noindent where $\phi=\phi(u)$ is an arbitrary function, $\phi'$,
$\phi''$ and $\phi'''$ are its first, second and third
derivatives.

In this paper we will use the algebra \eqref{eq algebra} as an
example for consideration of two aspects related to
infinite-dimensional algebras whose basis operators include
arbitrary functions - description of their differential invariants
and investigation of structure of such algebras. For both aspects
we suggest using discretisation of basis operators of such
algebras.

Practical finding of differential invariants of
infinite-dimensional algebras was considered in many papers, e.g.
in a number of papers by N.H.~Ibragimov and his coauthors (see
e.g.~\cite{Yehorchenko:I,Yehorchenko:IS}). Finding of the
differential invariants of the algebra \eqref{eq algebra} was
considered in \cite{Yehorchenko:TTV2}, where such invariants were
used for characterisation of linearisable equations from the class
\eqref{wave class}.

However, the ad-hoc approaches used in earlier papers did not
allow full rigorous description of functional bases of such
differential invariants. The algorithm we proposed in
\cite{Yehorchenko:spt2004} gives such description that is
essential for applications.

Here we use this method for an equivalence algebra of a class of
the nonlinear wave equations. The obtained description of
functional bases of the differential invariants allows using them
for characterisation of equations - that is, for each pair of
equations from the class \eqref{wave class} we can determine
whether these equations are equivalent under transformations of
the type \eqref{eq-transf}. In the case of ODE it had been already
shown that knowledge of such invariants gives both necessary and
sufficient conditions of equivalence \cite{Yehorchenko:berth}.

\section{Differential Invariants: Background}

Investigation of differential invariants usually stems from the
problem of description of equations that are invariant under
certain algebras (see e.g. \cite{Yehorchenko:FY}). Equations
invariant under certain algebras can be presented as functions of
absolute differential invariants or as relative differential
invariants.

The theory related to searching of differential invariants of
finite-dimensional Lie algebras is based on the classical results
by S.~Lie \cite{Yehorchenko:LieDI}. The relevant definitions and
theorems can be found e.g. in
\cite{Yehorchenko:book1,Yehorchenko:Ovs-eng}.

We consider\/description of differential invariants for Lie
algebras {Q} that consist of infinitesimal operators of the form $
Q_i=\xi_i^s(x,u)\partial_{x_s}+\eta_i^r(x,u)\partial_{u^r}$. Here
$x=(x^1,x^2,\ldots,x^n)$, $u=(u^1,\ldots,u^m)$. We mean summation
over the repeated indices.

{\bf Definition 2.} {\it  The function \[ F = F(x,
u,\mathop{u}\limits_1,\ldots,\mathop{u}\limits_l),
\]
\noindent where $\mathop{u}\limits_k$ is the set of all
$k$th-order partial derivatives of the function $u$ is called
a~differential invariant for the Lie algebra $L$ with basis
elements $Q_i$ of the form (0.1) $(L=\langle Q_i\rangle)$ if it is
an invariant of the $(l-r)$-th prolongation of this algebra:
\begin{equation}
\mathop{Q}\limits^l\!{}_s F(x,
u,\mathop{u}\limits_1,\ldots,\mathop{u}\limits_l)= \lambda_s(x,
u,\mathop{u}\limits_1,\ldots,\mathop{u}\limits_l)F, \end{equation}
\noindent where the $\lambda_s$ are some functions; when
$\lambda_i=0$, $F$ is called an absolute invariant; when
$\lambda_i \ne 0$, it is a relative invariant.}

For the prolongation formulae for infinitesimal operators see
\cite{Yehorchenko:book1,Yehorchenko:Ovs-eng}.

We will identify the order of a differential invariant of an
equivalence algebra by the order of its highest derivative of any
of the arbitrary functions in the respective class of equations.

The order of the prolongation needed to find $l$th-order
differential invariants may be smaller than $l$, when coefficients
of the basis operators of the equivalence algebra include
derivatives of arbitrary functions in the class of equations.

When speaking of differential invariants we will always mean
absolute differential invariants.

{\bf Definition 3.} {\it A maximal set of functionally independent
invariants of order $l_1\leq l$ of the Lie algebra $L$ is called a
functional basis of the $l$th-order differential invariants for
the algebra $L$.}

All absolute invariants of a particular order can be presented as
functions of invariants from a functional basis. The number of
invariants (up to  a certain particular order~$r$) in a functional
basis  is determined as difference between the number of all
derivatives up to the $l$-th order and both dependent and
independent variables, and of the general rank of $l$-th Lie
prolongation of the basis operators of the algebra under
consideration.

The case of infinite-dimensional algebras appears more complicated
than a finite-dimensional one, as the bases of such algebras
contain infinite (countable) number of operators (e.g.\ Virasoro
and Kac-Moody algebras), and/or infinitesimal operators having
arbitrary functions as coefficients (e.g. the equivalence algebra
\eqref{eq algebra}).

However, it appears that despite the name ``infinite-dimensional''
and actually having infinite-dimensional bases, such algebras, for
a fixed rank of their Lie prolongations and for the purpose of
finding their differential invariants can be treated as
finite-dimensional. The ranks of $l$-th Lie prolongations of basis
operators appear to be finite for each fixed $l$. Though, unlike
finite-dimensional algebras, these ranks do not stabilise, or do
not reach any fixed value with increase of the prolongation rank.

Finiteness of such rank is discussed in \cite{Yehorchenko:munoz},
where finiteness of a functional basis of differential invariants
was proved.

\section{Discretisation of the basis operators containing
arbitrary functions}

Here we present a systematic procedure (see
\cite{Yehorchenko:spt2004}) that considerable simplifies
previously utilised calculations of differential invariants for
the infinite-dimensional equivalence algebras. Instead of
arbitrary functions in the coefficients of the basis operators we
use expansions of these functions into Taylor series.

Here we deal with arbitrary functions, and in principle it may not
be always possible to expand them into Taylor series. These
functions have to be infinitely differentiable (due to commutation
condition in the definition of the Lie algebra),  but that does
not mean that they are analytical. However, for the purpose of
calculation of differential invariants of infinite-dimensional
algebras we can reasonably limit ourselves with consideration of
only analytical functions in coefficients of basis operators, and
with finite number of such arbitrary functions in coefficients of
basis operators.

Expansion of coefficients into series allows replacement of
operators with arbitrary functions with infinite series of
infinitesimal operators without such arbitrary functions. That
allows straightforward calculation of the prolongations' ranks (in
some cases the rank is equal to the number of variables and
derivatives, and then it is easy to see without any further
calculations that there are no absolute invariants of the
respective order).

\noindent {\bf Statement.} For a fixed order $l$ there exists a
functional basis of any Lie algebra, including
infinite-dimensional algebras with finite number of such arbitrary
functions in coefficients of basis operators or with countable
infinite sequences of basis operators with no arbitrary functions.

For our example of an equivalence algebra we get the following
representation of the basis operators:

$$
Y_0=x\partial_t+t\partial_x, Y_1=\partial_t, Y_2=\partial_x,
$$
\begin{equation}
Y_3=x\partial_x+t\partial_t- 2f\partial_f- \sigma\partial_\sigma
-2f_u\partial_{f_u},\label{eq algebra k}
\end{equation}
$$
 Y_\phi^k=u^k\partial_u+2ku^{k-1}\sigma\partial_\sigma+2(ku^{k-1}f+k(k-1)u^{k-2}\sigma)\partial_f+
 $$
 $$
(k(k-1)u^{k-2}f+k(k-1)(k-2)u^{k-3})\sigma-2\sigma f_\sigma
k(k-1)u^{k-2})\partial_{f_u}+
$$
 $$
 (k(k-1)u^{k-2}-f_\sigma ku^{k-1})\partial_{f_\sigma},
 $$

\section{Structure of the Discretised Equivalence Algebra}

The commutation relations for basis operators of the algebra
\eqref{eq algebra k} are as follows:
$$
[Y_\phi^n,Y_\phi^m]=(m-n)Y_\phi^{m+n-1},
$$
$$
[Y_0,Y_\phi^k]=[Y_1,Y_\phi^k]=[Y_2,Y_\phi^k]=[Y_3,Y_\phi^k]=[Y_0,Y_3]=[Y_1,Y_2]=
0,
$$
$$
[Y_0,Y_1]=-Y_2, [Y_0,Y_2]=-Y_1, [Y_1,Y_3]=Y_1, [Y_2,Y_3]=Y_2.
$$

It is easy to see that the largest $k$ for which the set
$\{Y_0,Y_1,Y_2,Y_3,Y_\phi^k \}$ forms a finite-dimensional algebra
is equal to 2.

\section{Differential Invariants for an Equivalence Algebra}
Coefficients of the basis operators of the algebra \eqref{eq
algebra k} contain first derivatives of the arbitrary function
$f$, so the basis operators themselves would give first-order
differential invariants (if exist), and their first prolongations
will give second-order differential invariants.

To find differential invariants for the algebra with the
discretised basis operators \eqref{eq algebra k} we find the
so-called minimal generating set of operators, that is the minimal
finite set of operators from the set \eqref{eq algebra k} having
the same general rank as the whole set \eqref{eq algebra k}. The
number of invariants in the functional basis of the order $l$ is
equal to the difference of the number of variables entering these
differential invariants (dependent and independent variables, and
derivatives of the relevant order), and of the general rank of the
needed prolongation of the basis operators.

The basis operators include first derivatives of the function $f$
and may in principle give first-order differential invariants.
However, we have 7 variables $(t,x,u,\sigma,f,f_u,f_\sigma)$, and
the general rank of the operators \eqref{eq algebra k} is equal to
7. We have no absolute differential invariants of the first order,
but one relative differential invariant
\begin{equation}
R=\sigma f_\sigma-f, \label{inv1}
\end{equation}
\noindent representing a special manifold for the algebra
\eqref{eq algebra k} where its rank is equal to 6.

A functional basis of the second-order absolute differential
invariants may be taken as

\begin{equation}
R_1=\frac{\sigma f_{\sigma\sigma}}{\sigma f_\sigma-f},
R_2=\frac{-2\sigma^2ff_{\sigma\sigma}+\sigma(f_u-\sigma f_{\sigma
u})+f(\sigma f_\sigma-f)}{(\sigma f_\sigma-f)^2} \label{inv2}
\end{equation}

\noindent (with $\sigma f_\sigma-f \ne 0$).

The invariants \eqref{inv2} were found in \cite{Yehorchenko:TTV2}.
Our approach allows proving that they form a functional basis of
the second-order absolute differential invariants and using them
for characterisation of the equations from the class \eqref{wave
class}.

We will be looking for invariants in the form
$F=F(t,x,u,\sigma,f,f_u,f_\sigma,f_{uu},f_{u\sigma},f_{\sigma\sigma})$
- depending on 10 variables. Here invariants depending on the
second derivatives of $u$ may be separated, and we do not need
them for our purpose of characterisation of the equations.

We will not write down the first prolongation of the operators
\eqref{eq algebra k} as it is quite cumbersome, but by direct
computation is it possible to determine that it general rank is
equal to 8, and as we have 10 variables, a functional basis will
contain 10-8=2 invariants. It is also easy to check that the
invariants \eqref{inv2} are functionally independent, so they
indeed form a functional basis.

The steps for calculation of functional bases of absolute
differential invariants for infinite-dimensional algebras with
arbitrary functions in basis operators are as follows:
\begin{enumerate}
\itemsep=0pt

\item[1.] Expand the arbitrary functions in the coefficients of
the operators into Taylor series.

\item[2.] Transform the set of operators with arbitrary functions
into discrete infinite set without arbitrary functions.

\item[3.] Find needed prolongations of the basis operators of the
equivalence algebra.

\item[4.] Calculate rank of the prolongation of the algebra.

\item[5.] Find a minimal ``generating set'' of operators with the
rank of their prolongation equal to that of the prolongation of
the algebra.

\item[6.]  Find a functional basis using the ``generating set''.
\end{enumerate}

\section{Characterisation of the differential equations}

In our study of characterisation of partial differential equations
we will follow the ideas presented in \cite{Yehorchenko:berth}
where a similar problem was considered for ODE.

For particular forms of equations of the form \eqref{wave class}
values of the invariants of the equivalence algebra will be some
functions of $u$ and $\sigma$ -
$R_1(u,\sigma,f)=\rho_1(u,\sigma)$,
$R_2(u,\sigma,f)=\rho_2(u,\sigma)$. We can prove that values of
all higher order differential invariants can be presented as
functions of $\rho_1(u,\sigma)$, $\rho_2(u,\sigma)$ (that means
that they form a fundamental basis for such algebra), and thus
values of these invariants will determine equivalence classes for
the class of equations \eqref{wave class}.

{\bf Statement}. Two equations
\begin{equation} \label{wave1}
u_{tt} - u_{xx}=f_1(u, \sigma),
\end{equation}
\noindent and
\begin{equation} \label{wave2}
u_{tt} - u_{xx}=f_2(u, \sigma),
\end{equation}
\noindent  are equivalent if and only if the values of the
differential invariants \eqref{inv2} for these equations are the
same:
$$
R_1(u,\sigma,f_1)=R_1(u,\sigma,f_2),\ \
R_2(u,\sigma,f_1)=R_2(u,\sigma,f_2).
$$

Treating the expressions
$$
R_1=\frac{\sigma f_{\sigma\sigma}}{\sigma
f_\sigma-f}=\rho_1(u,\sigma), $$
\begin{equation}
R_2=\frac{-2\sigma^2ff_{\sigma\sigma}+\sigma(f_u-\sigma f_{\sigma
u})+f(\sigma f_\sigma-f)}{(\sigma f_\sigma-f)^2}=\rho_2(u,\sigma)
\label{inv2a}
\end{equation}
\noindent as the system of partial differential equations on the
function $f$, it is possible to describe all equations in a
particular equivalence class.

\section{Conclusions}

We presented an approach for investigation of infinite-dimensional
algebras of the first-order differential operators on a particular
example of an equivalence algebra of a class of PDE \eqref{wave
class}. The presented approach for finding differential invariants
of infinite-dimensional algebras allows characterisation of
equivalent differential equations and description of equations
invariant under certain infinite-dimensional algebras.

For the class of equations \eqref{wave class} we have found the
fundamental basis of the second-order differential invariants of
its equivalence algebra and showed that knowing these invariants
is sufficient for characterisation of equations from this class,
that is to tell whether two equations from the class \eqref{wave
class} are equivalent.

This approach is very promising for characterisation of
equivalence classes represented by many interesting equations, and
description all equations equivalent (reducible by local
transformations) to specific equations.

%


\section*{Acknowledgments}

I would like to thank the organisers of the VII International
Workshop "Lie Theory and Its Applications in Physics" for the
wonderful conference and for financial support to attend the
conference.


\end{document}